\documentclass[aip,jcp, reprint, amssymb, amsmath,numerical, superscriptaddress]{revtex4-1}
\usepackage{graphicx}%
\usepackage{color}
\usepackage{graphics}
\usepackage{epstopdf}
\usepackage{dcolumn}%
\usepackage{bm}%

\usepackage{amsmath, amssymb, amsthm}

\newcommand{\be}{\begin{equation}}
\newcommand{\ee}{\end{equation}}
\newcommand{\bea}{\begin{eqnarray}}
\newcommand{\eea}{\end{eqnarray}}

\def\MnCO{Mn$_2$(CO)$_{10}\;$}
\def\cm{cm$^{-1}\,$}

\begin{document}
%\onecolumn

\title
{Carbonyl Vibrational Wave Packet Circulation in \MnCO Driven by Ultrashort Polarized Laser Pulses}
\author{Mahmoud K. Abdel-Latif}

\affiliation{Institut f\"{u}r Physik, Universit\"{a}t Rostock, D-18051 Rostock, Germany}
\affiliation{Chemistry Department, Faculty of Science, Beni-Suef University, Beni-Suef, Egypt}

\author{Oliver K\"uhn}
\email{oliver.kuehn@uni-rostock.de }
\affiliation{Institut f\"{u}r Physik, Universit\"{a}t Rostock, D-18051 Rostock, Germany}

\begin{abstract}
The excitation of the degenerate $E_1$ carbonyl stretching vibrations in dimanganese decacarbonyl is shown to trigger wave packet circulation in the subspace of these two modes. On the time scale of about 5 picoseconds intramolecular anharmonic couplings do not cause appreciable disturbance, even under conditions where the two $E_1$ modes are excited by up to about two vibrational quanta each. The compactness of the circulating wave packet is shown to depend strongly on the excitation conditions such as pulse duration and field strength. Numerical results for the solution of the seven-dimensional vibrational Schr\"odinger equation are obtained for a density functional theory based potential energy surface and using the multi-configuration time-dependent Hartree method.\end{abstract}
%\pacs{???}
%\keywords{quantum dynamics, laser control, MCTDH, organometallics}
\date{\today}%
\maketitle

\section{Introduction}
\label{sec:intro}
%%%%%%%%%%%%%%%%%%%%%%%%%%%%%%%%%%%%%%%
%
Recently, considerable attention has been paid to the excitation of degenerate electronic and vibrational states using 
circularly polarized laser light. Manz and coworkers have been the first to focus on the possibility of the excitation of  electronic ring currents by laser-triggered formation of time-dependent hybrid states, e.g. in a Mg-porphyrin model \cite{barth06:2962,barth06:2400} as well as in the diatomic AlCl. \cite{barth08:263} A particularly interesting aspect has been the accompanying generation of giant magnetic fields. \cite{barth07:012510} Subsequently, this work was extended to the vibrational domain in the electronic ground state. Here, degenerate bending vibrations of linear triatomic species such as $^{114}$CdH$_2$ \cite{barth08:89} and FHF$^-\;$ \cite{barth08:1311} have been excited to yield unidirectional pseudorotations. In contrast to previous work on the pseudorotation of Na$_3$ in the electronic $B$ state excited by linearly polarized light, \cite{gaus93:12509} unidirectionality is achieved by circular polarization. 

The present contribution is motivated by the question whether the dynamics of degenerate vibrations can be driven in more complex molecules, where one has to compete with intramolecular vibrational energy redistribution. \cite{may11} Promising candidates are metal-carbonyl compounds due to the strong and spectrally distinct infrared (IR) absorption of the carbonyl stretching vibrations. Previous studies of metal-carbonyl compounds have been focused especially on laser-driven CO vibrational ladder climbing. \cite{arrivo95:247,witte03:2021,ventalon04:13216} A particularly interesting example has been  the experimental demonstration of ladder climbing up to the $v_{\rm CO}=6$ state in carboxyhemoglobin by Joffre and coworkers. \cite{ventalon04:13216} For this example, Meier and Heitz have developed a non-reactive gas phase model to simulate the excitation dynamics by using local control theory. \cite{meier05:044504} It should be noted, however, that the main goal of the experiment had been to achieve controlled metal-carbonyl bond breaking assisted by anharmonic couplings between the initially excited CO mode and the metal-CO bond. A quantum dynamics study employing a reactive model provided support for the experimental finding that this anharmonic coupling is too weak to cause bond breaking on the time scale of a few picoseconds.\cite{kuhn05:48,kuhn09:329} The lack of sufficient anharmonic metal-CO coupling has also been found in the related model study of MnBr(CO)$_5$ using a three-dimensional reactive Hamiltonian. \cite{gollub07:369} Further experimental progress in vibrational ladder climbing has been reported in Ref. \citenum{strasfeld07:038102} where an evolutionary algorithm was used for pulse shaping such as to highly excite a CO mode in W(CO)$_6$. 

\begin{figure}
\begin{center}
\includegraphics[width=0.52\textwidth]{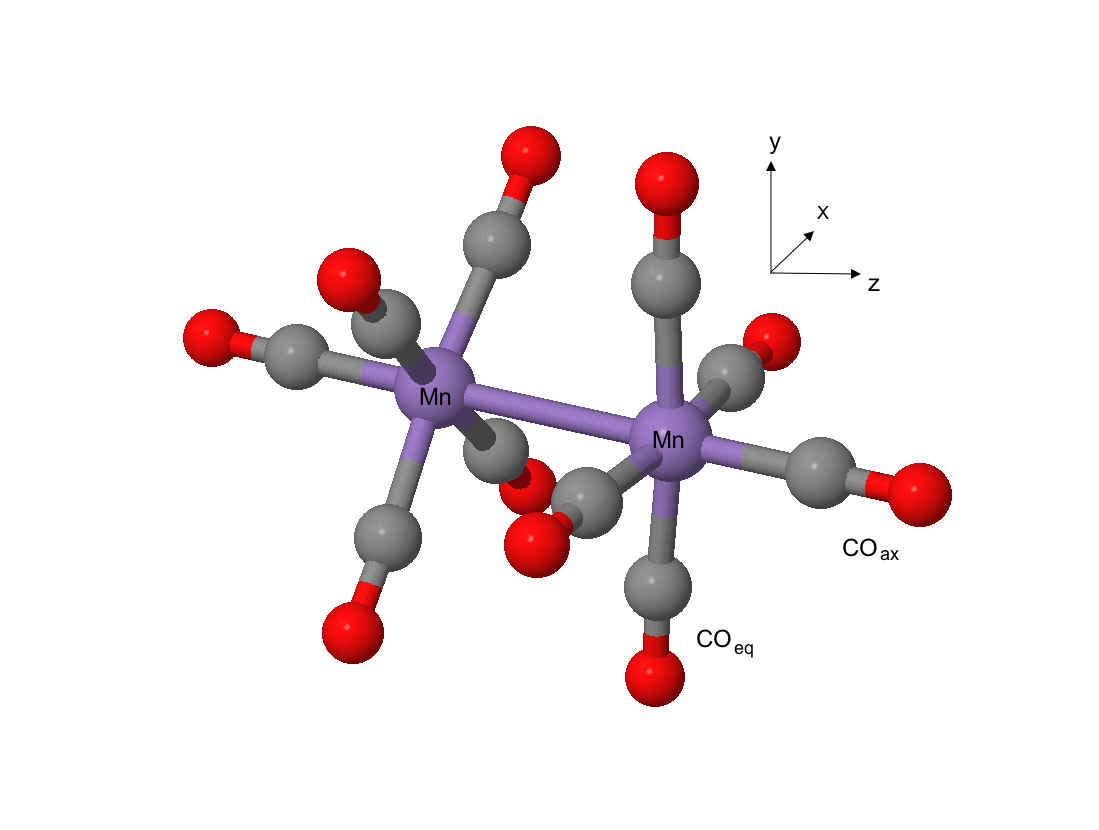}
\caption{Optimized D$_{\rm 4d}$ structure of \MnCO using DFT/BP86 with a TZVP basis set.}
\label{fig:1}
\end{center}
\end{figure}

In the following we will consider dimanganese decacarbonyl \MnCO (cf. Fig. \ref{fig:1}) as a model system. In the past it was in particular the photochemistry of \MnCO that attracted most attention (see, e.g., Ref. \citenum{kuhn00:199} and references cited therein). More recently, the prospect of having ten coupled carbonyl vibrations attracted the interest of two-dimensional IR spectroscopists. \cite{nee08:084503,baiz09:2433,baiz09:1395,king10:10590} This included in particular a detailed investigation of the coupled CO fundamental and low-order overtone and combination tone vibrations using  anharmonic vibrational perturbation theory for the subset of CO modes, which were treated by anharmonic couplings up to the fourth order. \cite{baiz09:2433} The anharmonic level shifts for all  CO vibrations were found to be below 10 \cm. 

In Section \ref{sec:eq} we will present results on  the quantum chemical characterization of the global minimum structure of the \MnCO model system. Based on this structure we will develop a seven-dimensional (7D) model Hamiltonian in Section \ref{sec:model}, which is designed such as to describe the dynamics of the two degenerate $E_1$ CO vibrational normal modes and their anharmonic coupling to the remaining degrees of freedom, which will be called a bath. Further, we will comment on the numerical solution of the time-dependent Schr\"odinger equation, which is done using the  multiconfiguration time-dependent Hartree (MCTDH) approach to wave packet dynamics. \cite{meyer90:73,beck00:1} Numerical results on the laser-driven dynamics  will be discussed in Section \ref{sec:res} and conclusions are present in Section \ref{sec:conc}.
%%%%%%%%%%%%%%%%%%%%%%%%%%%%%%%%%%%%%%%
\section{Theory}
\subsection{Characterization of the Equilibrium Geometry}
\label{sec:eq}
%%%%%%%%%%%%%%%%%%%%%%%%%%%%%%%%%%%%%%%
%
The structure of \MnCO (cf. Fig. \ref{fig:1}) has been optimized under D$_{\rm 4d}$ symmetry constraint employing Density Functional Theory (DFT) with the BP86 functional and a triple zeta (TZVP) basis set as implemented in the Gaussian 03 program package. \cite{gaussian03} Available crystal structure data are well reproduced as seen from the comparison given in Tab. I.
\begin{table}[b]
\begin{center}
\begin{tabular}{|c|cccc|}\hline  & calc. & Ref. \citenum{dahl63:419} & Ref. \citenum{almenningen69:685} & Ref. \citenum{martin82:6} \\\hline\hline 
Mn-Mn & 2.97 & 2.92 & 2.98 & 2.90 \\
\hline 
Mn-CO$_{\rm ax}$ & 1.81 & 1.73 & 1.80 & 1.82 \\
\hline Mn-CO$_{\rm eq}$ & 1.85 & 1.83 & 1.87 & 1.86 \\
\hline C-O$_{\rm ax}$ & 1.16 & 1.16 & 1.15 & 1.15 \\
\hline C-O$_{\rm eq}$  & 1.15 & 1.16 & 1.15 & 1.14 \\
\hline Mn-Mn-CO$_{\rm eq}$ & 86.8 & 86.2 & 86.6 & 86.1 \\
\hline\end{tabular} 
\caption{Comparison between calculated geometrical parameters of the D$_{\rm 4d}$ stationary point of \MnCO and crystal structure data (bond lengths in \AA, bond angles in degrees).}
\end{center}
\label{tab-geo}
\end{table}

The stationary points were further characterized by a normal mode frequency analysis. The results for the IR active CO stretching vibrations are summarized in Tab. II. Overall, the agreement with experiment is quite reasonable even at the level of the harmonic approximation. Moreover, comparison with Ref.  \citenum{baiz09:2433} indicates, that there
is only a small dependence on the basis set (double vs. triple zeta).

\begin{table}[b]
\begin{center}\begin{tabular}{|c|c|cccc|}
\hline 
mode & polariz. &calc. & exp. \cite{haas67:2996} & exp. \cite{parker83:463} & calc. \cite{baiz09:2433} \\
\hline\hline  
B$_2$ & z & 1988 (788) & 1983 & 1992 & 1981 \\
\hline 
E$_1$ &(x,y) & 2014 (2224) & 2014 & 2025 & 2005 \\
\hline 
B$_2$ & z &2045 (1348) & 2045 & 2053 & 2036 \\
\hline 
\end{tabular} 
\caption{Comparison between calculated (harmonic, not scaled) and measured (Ref. \citenum{parker83:463} in gas phase, Ref.  \citenum{haas67:2996} in $n$-hexane) frequencies of IR active carbonyl stretching vibrations (in \cm). Calculated IR intensities are given in parenthesis (in km/mol).}
\end{center}
\label{tab:ir}
\end{table}
\subsection{Model Hamiltonian and Equations of Motion}
\label{sec:model}
%%%%%%%%%%%%%%%%%%%%%%%%%%%%%%%%%%%%%%%
%
The potential energy surface has been expressed in terms of normal mode coordinates, i.e. $V=V(\mathbf{ Q},\mathbf{ q})$. Thereby, the two normal mode displacement coordinates describing the degenerate E$_1$ modes  $\mathbf{Q}=(Q_1,Q_2)$, which are treated beyond  harmonic approximation, have been separated from the remaining $3N-8$ harmonic bath modes $\mathbf{ q}$. The coupling, $V_{\rm c}(\mathbf{ Q},\mathbf{ q})$, between both sets of coordinates  is modeled in linear order with respect to $\mathbf{ q}$. Neglecting any contributions from global rotations the molecular Hamiltonian becomes
\begin{equation}
\label{eq:ham_mol}
H_{\rm mol}= H_{\mathbf{ Q}} + H_{\mathbf{ q}} + V_{\rm c}(\mathbf{ Q},\mathbf{ q}),
\end{equation}
with the part describing the anharmonic coordinates
\begin{equation}
\label{ }
 H_{\mathbf{ Q}} = \frac{1}{2}(P_1^2 +P_2^2) + V(\mathbf{ Q},\mathbf{ q}=0), 
\end{equation}
the harmonic bath part
\begin{equation}
\label{ }
 H_{\mathbf{ q}} = \frac{1}{2} \sum_i \left( p_i^2+\omega_i^2 q_i^2\right),
\end{equation}
and the interaction
\begin{equation}
\label{eq:vc}
V_{\rm c}(\mathbf{ Q},\mathbf{ q}) = \frac{\partial V(\mathbf{ Q},\mathbf{ q}=\mathbf{0})}{\partial \mathbf{ q}} \mathbf{ q} = -\mathbf{ F}(\mathbf{ Q}) \mathbf{ q} \; .
\end{equation}
For \MnCO there are 58 bath modes, but not all of them will couple appreciably to the $\mathbf{ Q}$ modes. 
Using the forces $\mathbf{ F}(\mathbf{ Q}) $ the  reorganization energy  has been calculated according to the assumption that the frequencies do not change, i.e. $E^{\rm (reorg)}_i= [F_i(\mathbf{ Q})/\omega_i]^2$. \cite{may11} This quantity has been averaged on the $\mathbf{ Q}$ grid and the result is used to select the most relevant bath modes. The cut-off for the selection has been chosen such that modes whose averaged reorganization energy is substantially smaller than their harmonic frequencies are neglected. This gave a total of five relevant bath modes, which are combined with the anharmonic coordinates to a seven-dimensional model. The bath mode parameters  are compiled in Tab. \ref{tab:modes}. Notice that the average reorganization energy depends, of course, on the grid size. In the table we report the average over the full grid as well as over a reduced grid onto which the wave packet will be mostly localized. The major effect on the dynamics of the $\mathbf{ Q}$ modes will come from the coupling to the other CO vibrations, not only because the forces are the largest, but also because of the near resonance. Among the lower-frequency modes only $q_1$ couples appreciably; it is a breathing mode of all equatorial CO groups with respect to the two Mn centers.

\begin{table}[t]
\centering
\caption{Most important bath modes according the averaged reorganization energy $\langle E^{\rm (reorg)}\rangle$ (average over the full $\mathbf{ Q}$ grid and - in parentheses - over a reduced grid from -1 to 1  $a_0$(a.m.u.)$^{1/2}$). The parameters of the MCTDH implementation ($N_{\rm DVR}$: number of DVR points, $N_{\rm SPF}$: number of single particle functions) are given as well. Notice that modes $q_2$ and $q_5$ as well as $q_3$ and $q_4$ have been combined.}
\begin{tabular}[t]{|c|c|c|c|c|c|}
\hline
mode & $\omega_i$& $\langle E^{\rm (reorg)}\rangle$ & grid  & $N_{\rm DVR}$ & $N_{\rm SPF}$ \\
 &  (cm$^{-1}$) &  (cm$^{-1}$) & ($a_0$(a.m.u.)$^{1/2}$) &&\\
\hline
$q_1$ (A$_1$)    & 425   & 468 (72)  & -7.5:7.5 & 45 & 10 \\ 
$q_2$ (A$_1$)    & 1999 & 822  (122)& -1.7:1.7 & 25 & 10 \\ 
$q_3$ (E$_2$)    & 2016 & 1426   (204) & -1.7:1.7 & 25 & 10 \\
$q_4$ (E$_2$)    & 2016 & 1649  (239)& -1.7:1.7 & 25 & 10 \\
$q_5$ (A$_1$)    & 2106 & 4646    (692) & -1.7:1.7 & 25 & 10 \\
\hline
\end{tabular}
\label{tab:modes}
\end{table}
The molecular Hamiltonian is supplemented by the molecule-field interaction that is described in dipole approximation
\begin{equation}
\label{eq:field}
H_{\rm field}(t)= - \mathbf{ d}(\mathbf{ Q}) \mathbf{ E}(t),
\end{equation}
where the dipole moment surfaces for the different polarization directions have been obtained along with the potential energy surface for the selected anharmonic degrees of freedom. Notice that within the approximation given by Eq. (\ref{eq:vc}) there is no coupling to the other two IR active carbonyl modes (cf. Tab. II). Moreover, the latter are polarized in $z$-direction and will not be directly excited by the circularly polarized laser pulse. For the dynamics simulations it will be assumed that the molecule has been pre-oriented with the coordinate system as indicated in Fig. \ref{fig:1}. The laser field will be taken to be of the form
\bea
\label{eq:laser}
\mathbf{ E}(t) &=&\left(\begin{array}{c}
     E_x(t)\\
     E_y(t)  \\
      0 
\end{array}\right) \nonumber\\
&= &
 E_0 \Theta(t)\Theta(t_{\rm p}-t)\sin^2(t \pi/t_{\rm p}) 
\left(\begin{array}{c}
     \cos(\omega t)    \\
      \sin(\omega t)  \\
      0 
\end{array}\right).
\eea
\begin{figure*}
\begin{center}
\includegraphics[width=0.9\textwidth]{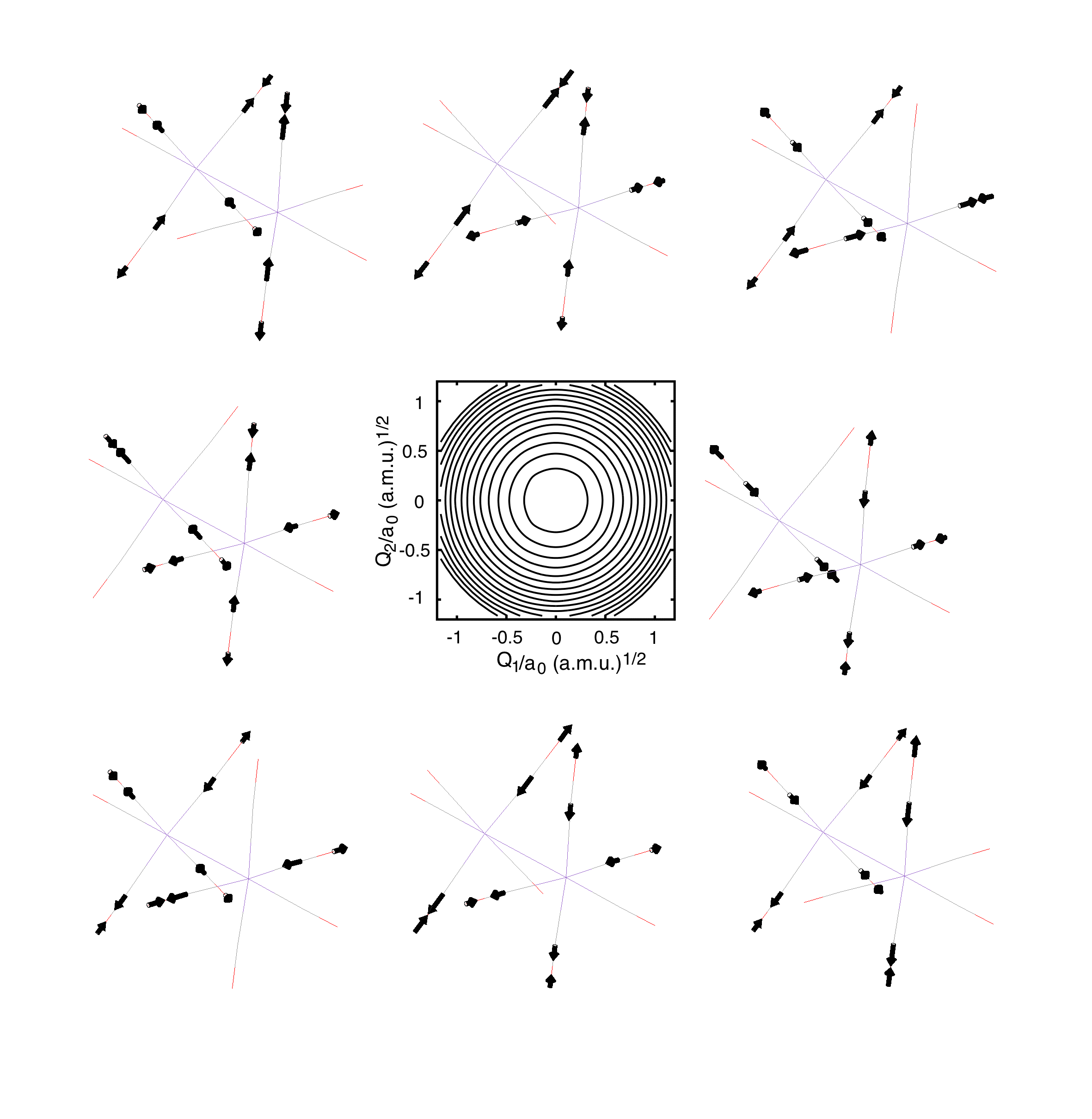}
\caption{Classical picture of vibrational wave packet circulation triggered by excitation of the two degenerate $E_1$ modes by means of a circularly polarized laser field. Shown are the displacement vectors for the superposition of the modes along a unit circle in steps of 45$^\circ$. In the central panel the two-dimensional potential $V(\mathbf{ Q},\mathbf{ q}=0)$ is shown (contour lines in steps of 2000 \cm starting at 2000 \cm).}
\label{fig:2}
\end{center}
\end{figure*}
Here $t_{\rm p}$ is the total pulse duration, $\omega$ the carrier frequency, and $E_0$ the field strength. These parameters will be chosen such as to achieve the excitation of a vibrational superposition state with respect to the two $E_1$ modes, which corresponds to a wave packet performing a clockwise circulation on the potential energy surface $V(\mathbf{ Q},\mathbf{ q}=0)$. The respective normal mode displacements are shown in Fig. \ref{fig:2}. Close inspection of this figure shows that during this circulation, the antisymmetric combination of stretching motions of equatorial CO groups which  are opposite to each other "moves" around the symmetry ($z$) axis. Of course, this classical picture applies only as long as the wave packet stays compact and an important question to be addressed will be the regime of validity of this classical analogue.

The 7D time-dependent Schr\"odinger equation
\begin{equation}
\label{eq:sgl}
i\hbar \frac{\partial}{\partial t} \Psi(\mathbf{ Q},\mathbf{ q};t) = [H_{\rm mol}+H_{\rm field}(t)] \Psi(\mathbf{ Q},\mathbf{ q};t)
\end{equation}
has been solved using the MCTDH approach \cite{meyer90:73,beck00:1} as implemented in the Heidelberg program package. \cite{mctdh84} The seven-dimensional wave function $\Psi(\mathbf{ Q},\mathbf{ q};t)$ is represented on a grid in terms of a harmonic oscillator discrete variable representation (DVR). A total of 25 DVR points on the grid $[-1.7:1.7]a_0$(a.m.u.)$^{1/2}$ together with 12 single particle functions (SPFs) are used for each of the $\mathbf{ Q}$ modes which are treated as a combined MCTDH particle. The parameters for the bath modes are compiled in Tab. \ref{tab:modes}. Potential energy and dipole moment surfaces have been fitted on this grid using the POTFIT algorithm. \cite{beck00:1}  Using this set of SPFs the natural orbital populations have been below $10^{-5}$.  The actual propagation is performed using the  variable mean field  scheme  with a 6th order Adams-Bashforth-Moulton integrator. The vibrational ground state, $|\Psi_0 \rangle$, has been calculated using the improved relaxation scheme \cite{meyer06:179}; some uncoupled (one-dimensional zero-order) states in the  $\mathbf{ Q}$-subspace will be used for the discussion of the dynamics.
%
%sssssssssssssssssssssssssssssssssssssssssssssssssss
\section{Results and Discussion}
\label{sec:res}
%sssssssssssssssssssssssssssssssssssssssssssssssssss
%
In the following we will present results for the time-dependent wave packet dynamics which have been obtained for a carrier frequency $\omega$, fixed to the fundamental transition of the $E_1$ modes, which is at 2014 \cm. The total propagation time has been set to 5 ps. First, we will discuss the effect of changing the field strength $E_0$, before the variation of the pulse length is addressed. A global analysis of the dynamics will be given in terms of the following quantities: (i) the energy absorbed by the molecule at the end of the pulse: $\Delta E _{\rm mol}=\langle \Psi(t_{\rm p})| H_{\rm mol} | \Psi(t_{\rm p})\rangle$-$\langle \Psi_0| H_{\rm mol} | \Psi_0\rangle$, (ii) the maximum energy change of the bath modes reached during the considered time interval of 5 ps: $\Delta E _{\rm bath}^{\rm (max)}$=max($\langle \Psi(t)| H_{\mathbf q} | \Psi(t)\rangle$-$\langle \Psi_0| H_{\mathbf q} | \Psi_0\rangle$), (iii) the maximum expectation value of the $E_1$ modes' coordinate operators during the 5 ps time interval: $Q^{\rm (max)}$=max($\langle \Psi(t)| {\mathbf Q} | \Psi(t)\rangle$), and (iv) the population of the ground state at the end of the pulse: $P_0=|\langle \Psi(t_{\rm p})|\Psi_0\rangle|^2$.
\subsection{Dependence on the Field Strength}  %------------------------------------------------------------------------------
\begin{table}
\centering
\caption{Analysis of the laser-driven wave packet dynamics for a 500 fs pulse which is resonant to the fundamental transition of the $E_1$ modes (cf. Eq. (\ref{eq:laser})).}
\begin{tabular}[t]{|c|c|c|c|c|}
\hline
$E_0$  & $\Delta E _{\rm mol}$ &$\Delta E _{\rm bath}^{\rm (max)}$ & $Q^{\rm (max)}$ & $P_0$ \\[0.5ex]
(mE$_{\rm h}/ea_0)$ &  (\cm) &  (\cm) & ($a_0\sqrt{\rm a.m.u}$) &\\
\hline
0.1     & 71 		& 6 		 & 0.05 & 0.97   \\ 
0.2      & 277 		& 18 	 & 0.09 & 0.87 \\ 
0.3     & 621 		& 62 	 & 0.13 & 0.73 \\ 
0.4     & 1099 	& 164  & 0.18 & 0.57 \\ 
0.5     & 1708 	& 341  & 0.22 & 0.42 \\ 
1.5     & 12733 & 4553 & 0.50 & 0.00  \\
\hline
\end{tabular}
\label{tab:E0}
\end{table}
A global analysis of the dynamics in dependence on the field strength and for a pulse duration of $t_{\rm p}=500$ fs is given in Tab. \ref{tab:E0}. Overall, we observe a monotonous decrease of the ground state population $P_0$ with increasing field amplitude. For the highest amplitude used, the ground state is completely depopulated. At the same time the absorbed energy $\Delta E _{\rm mol}$ increases, indicating vibrational ladder climbing. In fact for $E_0=1.5$ mE$_{\rm h}/ea_0$ the absorbed energy is compatible with an appreciable simultaneous population of the third vibrationally excited state in each of the two $\mathbf{ Q}$ modes. Analysis of the population shows that in fact many overtone and combination transitions are excited  with the $(\nu_1=2,\nu_2=2)$ state having the largest population. For the more moderate excitation with $E_0=0.5$ mE$_{\rm h}/ea_0$ the populations of the $\mathbf{ Q}$ modes' states is  about 20\% for $(0,1)$ and $(0,1)$, 8\% for $(1,1)$ and 4 \% for $(2,0)$ and $(0,2)$.  

\begin{figure}
\begin{center}
\includegraphics[width=0.45\textwidth]{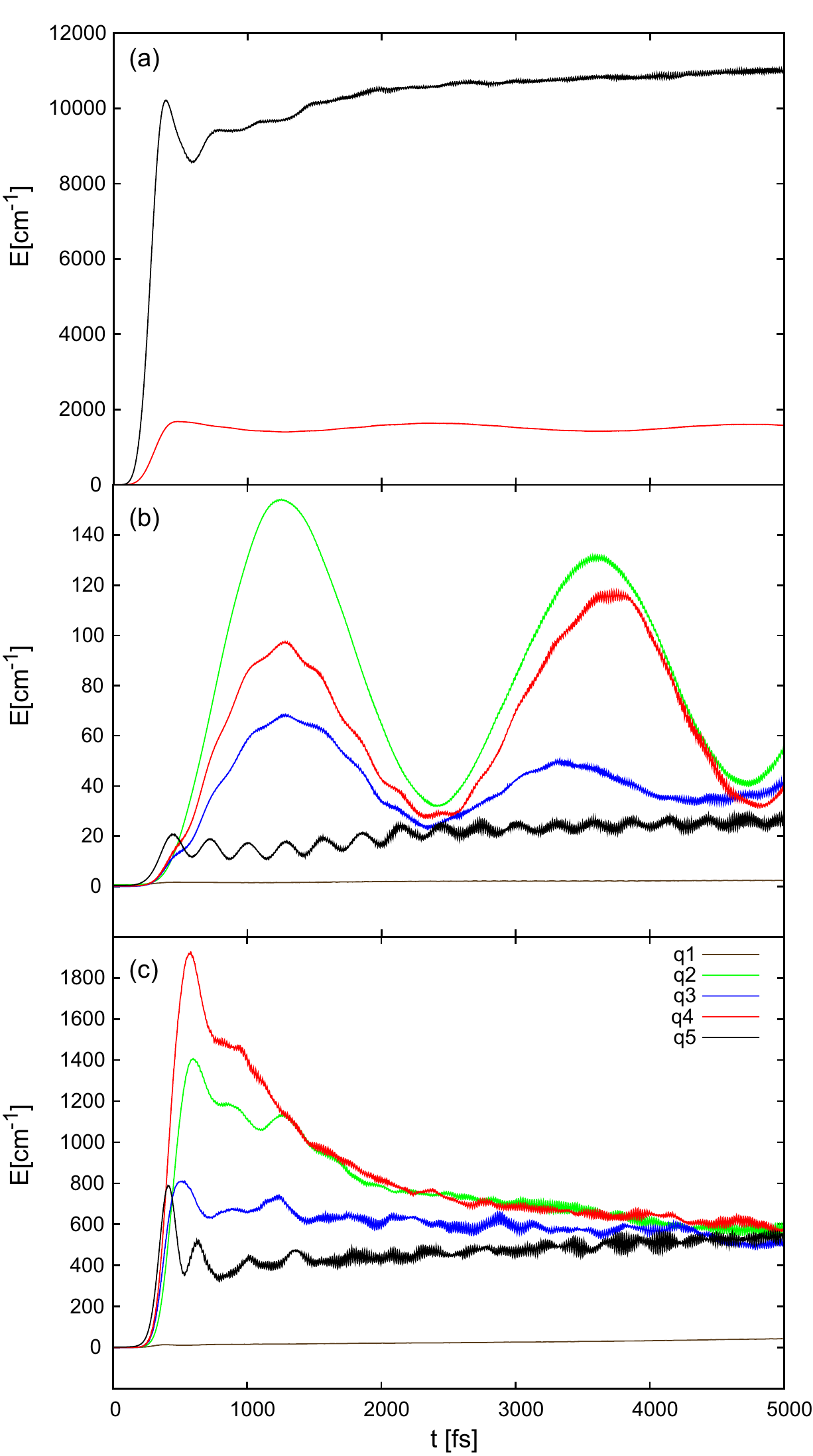}
\caption{(a) Expectation values of $H_\mathbf{ Q}$ for a 500 fs excitation with field strength $E_0=0.5$ mE$_{\rm h}/ea_0$ (lower curve)  and 1.5 mE$_{\rm h}/ea_0$ (upper curve). The respective expectation values of $H_\mathbf{ q}$ for the five bath modes are shown in panels (b) ($E_0=0.5$ mE$_{\rm h}/ea_0$) and (c) ($E_0=1.5$ mE$_{\rm h}/ea_0$). With increasing energy at about 1000 fs the curves correspond to $q_1$, $q_5$, $q_3$, $q_2$, and $q_4$ in panel (c), whereas in panel (b) $q_2$ and $q_4$ are interchanged (see also key in panel (c)).}
\label{fig:h0}
\end{center}
\end{figure}

The anharmonic coupling to the bath modes will increase along with  the absorbed energy, i.e. if the wave packet explores configurations away from equilibrium.  For moderate excitation conditions  the extent of anharmonic coupling is rather small, e.g. if the amount of absorbed energy corresponds to about one quantum of excitation in a CO mode ($E_0=0.5$ mE$_{\rm h}/ea_0$) we find  $\Delta E _{\rm mol}/\Delta E _{\rm bath}^{\rm (max)}\approx 5$. 
The effect of the bath becomes appreciable only under high excitation conditions, e.g., for $E_0=1.5$ mE$_{\rm h}/ea_0$ we find $\Delta E _{\rm mol}/\Delta E _{\rm bath}^{\rm (max)} \approx 2.8$. 
In Fig. \ref{fig:h0} we show the expectation values of the Hamiltonian operators for the system modes, $H_\mathbf{ Q}$, and the uncoupled bath modes, $H_\mathbf{ q}$, for two field strengths. As expected the energy of the  $\mathbf{ Q}$ modes rises during laser pulse excitation ($t_{\rm p}=500$ fs) as seen in panel (a). For the lower field strength the expectation value subsequently oscillates around the attained value signaling reversible energy exchange with the bath modes. Indeed the corresponding oscillations in the bath mode energies are seen in panel (b). Increasing the field strength, the energy exchange with the bath is more pronounced and instead of regular oscillations the onset of equilibration among the bath modes is observed in panel (c). Still there is no appreciable coupling to lower frequency modes that could lead to an energy flow out of the selected $\mathbf{ Q}$ modes. We notice from panels (b) and (c) that the low-frequency mode $q_1$, which was selected because it had the largest reorganization energy in the lower-frequency part of the spectrum is not appreciably excited.
Finally, we comment on the initial partitioning of energy among the bath modes. Inspecting Tab. III one might have anticipated that mode $q_5$ will dominate the bath dynamics. However, the fundamental frequency of the laser-excited $\mathbf{ Q}$ modes is 2014 \cm, which is resonant to $q_2-q_4$, but rather off-resonant with respect to $q_5$. Hence, energy exchange between the $\mathbf{ Q}$ and the $q_5$  modes is not very efficient.

\begin{figure}
\begin{center}
\includegraphics[width=0.5\textwidth]{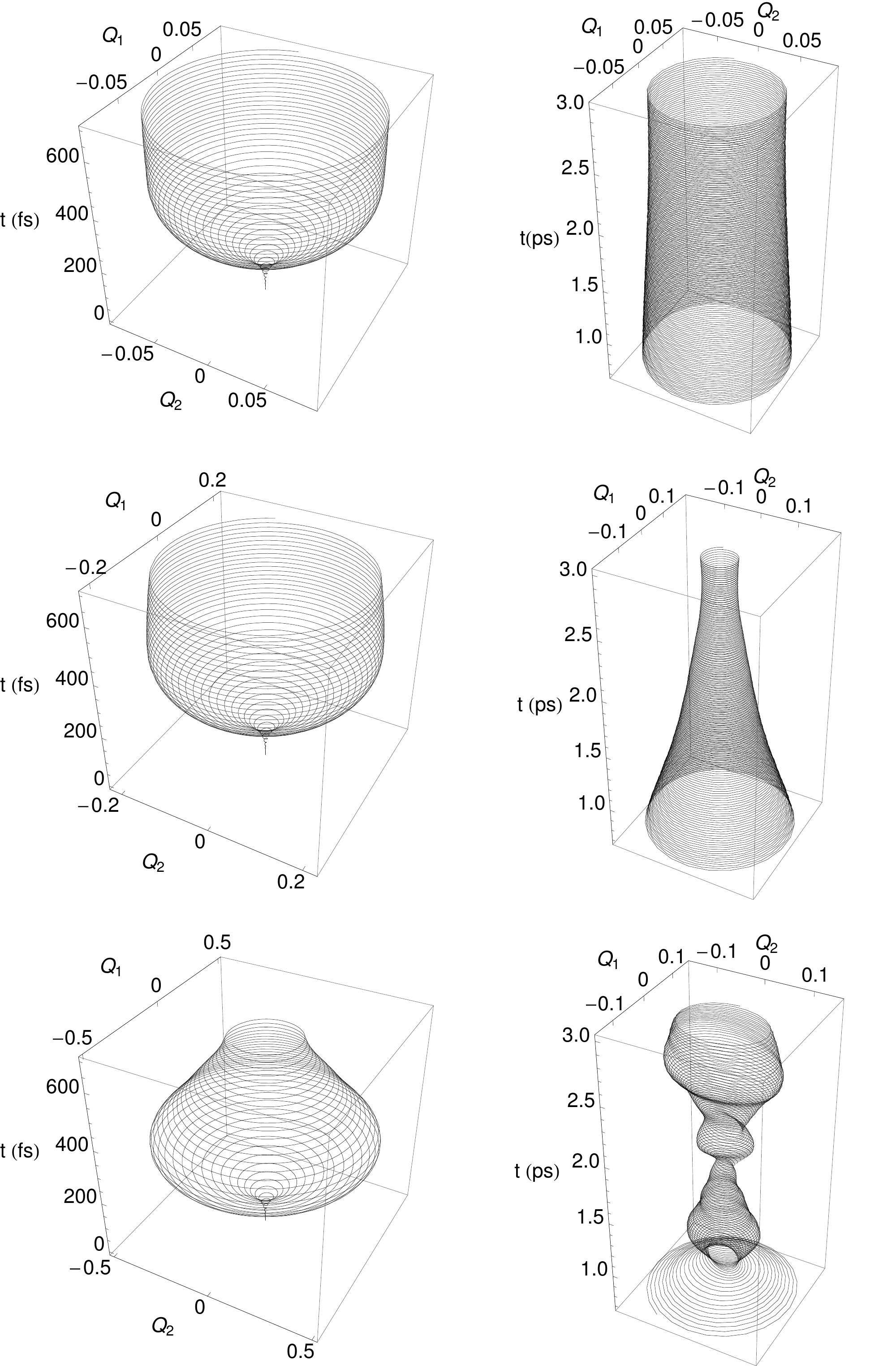}
\caption{Parametric plots of $(Q_1,Q_2)$ (in $a_0\sqrt{\rm a.m.u}$) for a 500 fs laser pulse excitation and a field strength $E_0$ (in mE$_{\rm h}/ea_0$) of 0.2 (upper row), 0.5 (middle row) and 1.5 (bottom row). (Notice the different axes scales).}
\label{fig:4}
\end{center}
\end{figure}

\begin{figure}
\begin{center}
\includegraphics[width=0.5\textwidth]{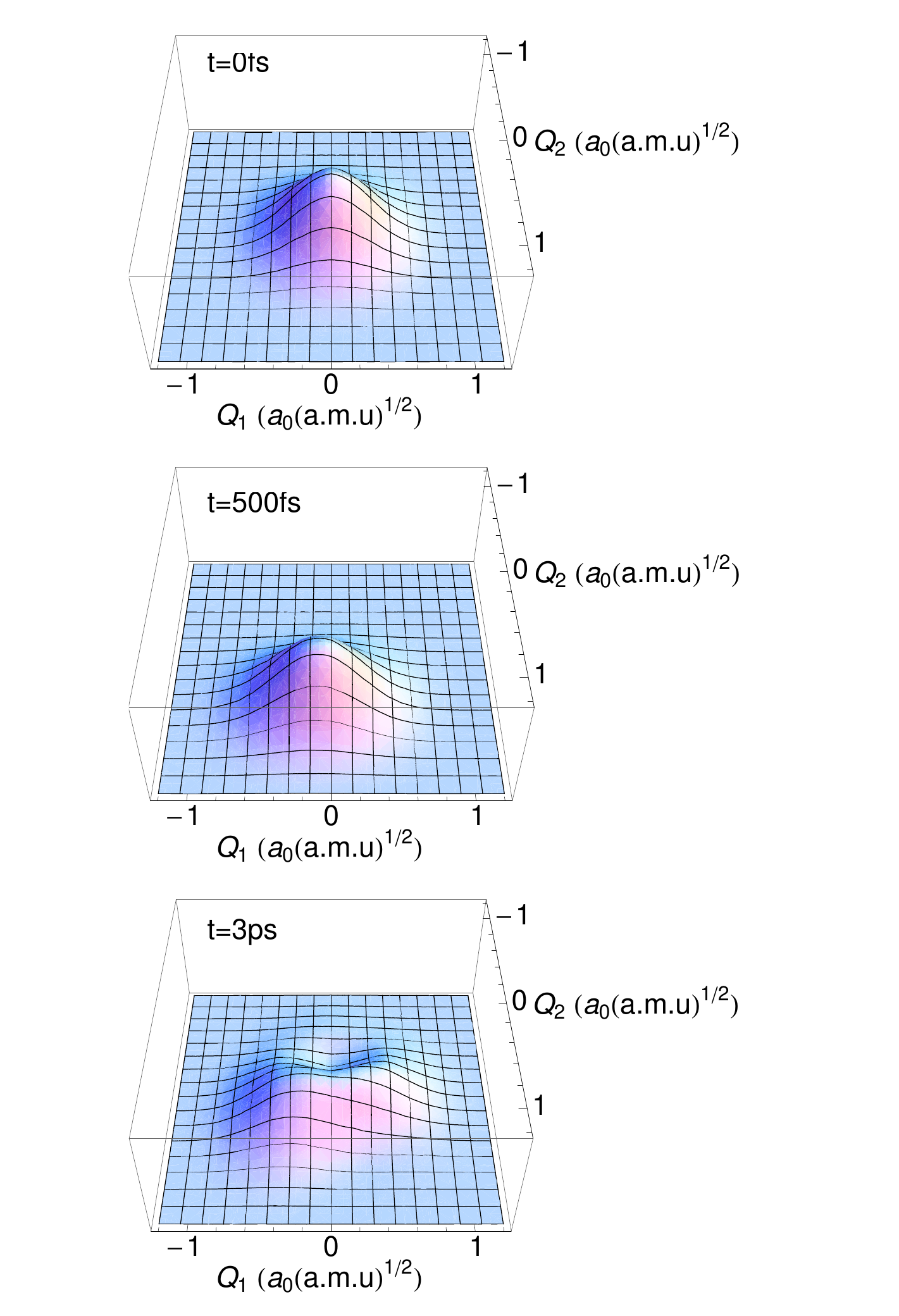}
\caption{Snapshots of the  time evolution of the reduced density with respect to the $\mathbf{ Q}$ modes for laser-driving with a pulse having the parameters $t_{\rm p}=500$ fs, $E_0=0.5$ mE$_{\rm h}/ea_0$, and $\omega/2\pi c=2014$ \cm.}
\label{fig:wp1}
\end{center}
\end{figure}

The focus of the present investigation is on the wave packet circulation with respect to the $E_1$ modes.  Inspecting Tab. \ref{tab:E0} we notice that the maximum bond elongation increases with the amount of absorbed energy. The value that is reached can be compared, e.g., with the variance of the ground state coordinate distribution which is 0.17 $a_0\sqrt{\rm a.m.u}$ for the $\mathbf{ Q}$ modes. Thus for the largest field strength we observe that the bond elongation is about three times this ground state variance. A more detailed picture is provided by inspecting the coordinate operator expectation values, $Q_i(t)=\langle \Psi(t)| Q_i | \Psi(t)\rangle$, which are shown in Fig. \ref{fig:4}. In the parametric plot the expectation values trace a clockwise circulating trajectory on the potential $V(\mathbf{ Q},\mathbf{ q}=\mathbf{ 0})$. For small field strengths the amplitude of this circulation increases during the excitation period of 500 fs (upper and middle panel of Fig. \ref{fig:4}). For the strongest field used (lower panel of Fig. \ref{fig:4}) the amplitude of the circulation starts to decrease already during the interaction with the pulse. Furthermore, after the excitation pulse is over the behavior shows a strong dependence on the field amplitude (right column of Fig. \ref{fig:4}). Judging from the amount of energy redistribution given in Tab. \ref{tab:E0} the decrease of $Q_i(t)$ can only in part be an effect of the anharmonic coupling to the bath.  Instead it is the wave packet dispersion that is mostly  responsible for this decrease of $Q_i(t)$. This is shown in  Fig. \ref{fig:wp1} where snapshots of the reduced density are plotted for the case of $E_0=0.5$ mE$_{\rm h}/ea_0$. 
One notices that the wave packet stays compact in the excitation period only, but then starts to disperse. Eventually, the wave packet is rather delocalized even though it is still circulating. However, due to the delocalization negative and positive coordinate values contribute to the expectation values along the $\mathbf{ Q}$ axes and therefore the expectation values become smaller and cannot reflect the wave packet circulation anymore. With increasing excitation more eigenstates contribute to the wave packet such that the dispersion effect becomes more pronounced and sets in already during the pulse.
\subsection{Dependence on the Pulse Duration}  %------------------------------------------------------------------------------
In the following we investigate the dependence of the laser-driven wave packet dynamics on the duration of the excitation pulse. As a reference case we have used $t_{\rm p}=500$ fs and $E_0=0.5$ mE$_{\rm h}/ea_0$. For a given duration $t_{\rm p}$ the field amplitudes have been chosen such as to keep the integrated field envelope at the same value as in the reference case. Note that we are not dealing with a two level system and that the pulses are no $\pi$-pulses. The pulse parameters are summarized in Tab. \ref{tab:tp}.
\begin{table}
\centering
\caption{Analysis of the laser-driven wave packet dynamics for pulses of different duration and amplitude. The integrated field envelope is identical in all cases.}
\begin{tabular}[t]{|c|c|c|c|c|c|}
\hline
$t_{\rm p}$ & $E_0$  & $\Delta E _{\rm mol}$ &$\Delta E _{\rm bath}^{\rm (max)}$  & $Q^{\rm (max)}$ & $P_0$ \\[0.5ex]
(fs) & (mE$_{\rm h}/ea_0)$ & (\cm)  & (\cm) & ($a_0\sqrt{\rm a.m.u}$) &\\
\hline
100       	& 2.5   	& 1812     	& 374 		& 0.23 		& 0.40 \\ 
500       	& 0.5   	& 1708     	& 341 		& 0.22   	& 0.42 \\ 
1000    		& 0.25 	& 1441  			& 273 		& 0.20   	& 0.47 \\ 
1500   		& 0.17 	& 1146     	& 205 		& 0.18 		& 0.54 \\ 
\hline
\end{tabular}
\label{tab:tp}
\end{table}

\begin{figure}
\begin{center}
\includegraphics[width=0.5\textwidth]{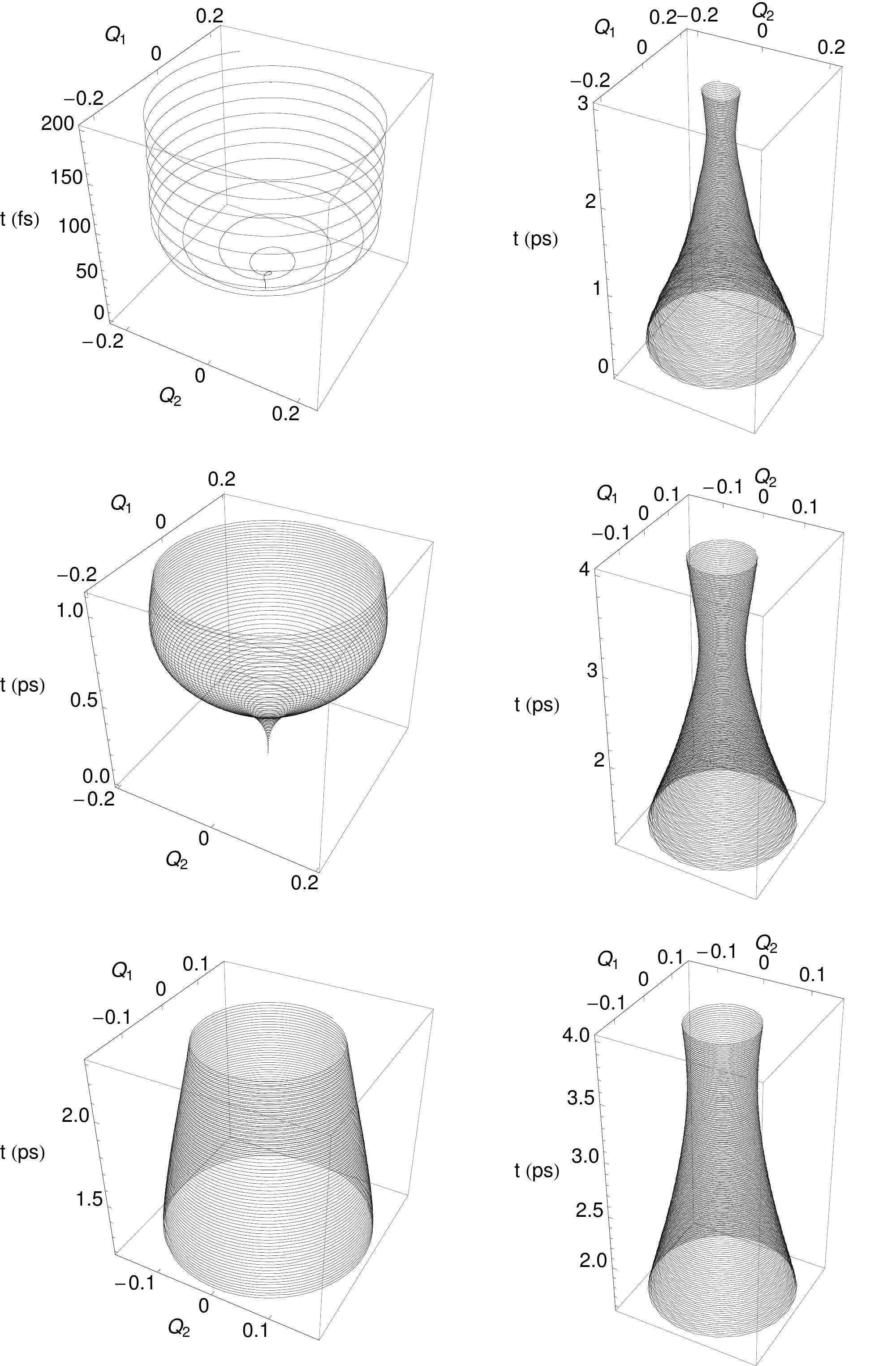}
\caption{Parametric plots of $(Q_1,Q_2)$ (in $a_0\sqrt{\rm a.m.u}$) for different pulse durations. Upper row: $t_{\rm p}=100$ fs, $E_0=2.5$ mE$_{\rm h}/ea_0$, middle row: $t_{\rm p}=1000$ fs, $E_0=0.25$ mE$_{\rm h}/ea_0$, bottom row: $t_{\rm p}=1500$ fs, $E_0=0.17$ mE$_{\rm h}/ea_0$. (Notice the different axes scales).}
\label{fig:tp}
\end{center}
\end{figure}

Table \ref{tab:tp} also contains the various expectation values, which give a global characterization of the dynamics. First, we notice that the different pulse durations give comparable depopulations of the vibrational ground state between 40 and  54\%. The amount of energy absorbed by the molecule decreases with the pulse duration and is largest for the shortest pulse ($t_{\rm p}=100$ fs), although the difference to the reference pulse is only 6\%. At the same time the excitation of bath modes decreases from a contribution to the total energy of 21\% for $t_{\rm p}=100$ fs down to 18\% for $t_{\rm p}=1500$. Moreover, the maximum radius of circulation of the expectation value of the $\mathbf{Q}$ coordinates is reduced slightly. 
Inspecting the time dependence of the $Q_i(t)$  in Fig. \ref{fig:tp} we notice that again we have an initial period, where the wave packet stays rather compact such that the coordinate expectation values give a reasonable picture of the moving wave packet. This is  followed by a period where the wave packet disperses such that the dynamics cannot be understood from the expectation values. 
As far as the pulse duration dependence is concerned the most striking observation from Fig. \ref{fig:tp} is that the time span of compact wave packet propagation can be  prolongated by increasing the parameter $t_{\rm p}$. 
In passing we note that this does also imply that the time span of circulation at about the full radius is longer, but of course not identical to $t_{\rm p}$. For $t_{\rm p}=500$ fs and $1500$ fs the expectation values $Q_i(t)$ are within 90\% of their maximum value during the interval $[450:800]$ fs and  $[950:1650]$ fs, respectively.
Since the anharmonic coupling is still small in the considered range of the potential energy surface corresponding to about one to two quanta of excitation of the $E_1$ modes, intramolecular energy redistribution doesn't destroy wave packet circulation even for the longest pulse duration of 1.5 ps and subsequent 3.5 ps free evolution.

%
%
%
%sssssssssssssssssssssssssssssssssssssssssssssssssss
\section{Summary}
\label{sec:conc}
%sssssssssssssssssssssssssssssssssssssssssssssssssss
%
We have demonstrated laser-driven wave packet circulation with respect to the two degenerate $E_1$ carbonyl stretching vibrations in a pre-oriented \MnCO model. The wave packet circulation corresponds to a vibrational excitation moving around the symmetry axis of the system. It has been shown that a classical picture holds approximately during the interaction with the excitation laser pulse. Later on, appreciable wave packet dispersion sets in such that the classical picture becomes meaningless, although the wave packet still circulates around the minimum of the potential energy surface. 

The possibility of wave packet circulation upon excitation of degenerate vibrational modes with circularly polarized laser light could have been anticipated from the work of J. Manz et al. on triatomic molecules. \cite{barth08:89,barth08:1311} Therefore, the most important result of the present study concerns the robustness of wave packet circulation with respect to intramolecular vibrational energy redistribution in a polyatomic molecule. To proof this we have considered the coupling of the two $E_1$ mode to other intramolecular degrees of freedom, which have been described in harmonic approximation. This resulted in a seven-dimensional model for which the time-dependent Schr\"odinger equation has been solved using the MCTDH approach. For moderate excitation conditions of one or two quanta in the $E_1$ modes, at most about 20 \% of the absorbed energy is disposed into the bath modes within a time span of 5 picoseconds. This finding appears to be encouraging and should stimulate experimental verification.
%
%sssssssssssssssssssssssssssssssssssssssssssssssssss
%
\section*{Acknowledgment}
We gratefully acknowledge financial support from the Deutsche Forschungsgemeinschaft (project Ku952/6-1).
%
%\bibliography{/Volumes/oliver.kuehn/Documents/Bibliothek/BibTeX/papers}

\end{document}